\documentclass[preprint]{aastex}

\shorttitle{Accretion of Terrestrial Planets in a Turbulent Disk}
\shortauthors{Ogihara et al.}

\begin{document}
  
\title{Accretion of Terrestrial Planets from Oligarchs \\
       in a Turbulent Disk}
\author{Masahiro Ogihara, Shigeru Ida} 
\affil{Department of Earth and Planetary Sciences,
Tokyo Institute of Technology,
Ookayama, Meguro-ku, Tokyo 152-8551, Japan}
\email{ogihara@geo.titech.ac.jp}

\and

\author{Alessandro Morbidelli} 
\affil{Observatoire de la C\^{o}te d'Azur,
CNRS UMR 6202, BP 4229, 06304 Nice Cedex 4, France}

\vspace{1cm} 

\section*{Abstract}
We have investigated the final accretion stage of terrestrial planets 
from Mars-mass protoplanets that formed through oligarchic growth
in a disk comparable to the minimum mass solar nebula (MMSN), 
through N-body simulation including random torques exerted by
disk turbulence due to Magneto-Rotational-Instability.
For the torques, we used the semi-analytical formula 
developed by \citet{lau04}.  
The damping of orbital eccentricities
(in all runs) and type-I migration (in some runs) due to
the tidal interactions with disk gas are also included.
Without any effect of disk gas,
Earth-mass planets are formed 
in terrestrial planet regions in a disk comparable to MMSN
but with too large orbital eccentricities
to be consistent with the present eccentricities of Earth and Venus
in our Solar system.  With the eccentricity damping caused by 
the tidal interaction with a remnant gas disk,
Earth-mass planets with 
eccentricities consistent with those of Earth and Venus
are formed in a limited range of disk gas surface density 
($\sim 10^{-4}$ times MMSN).
However, in this case, on average,
too many ($\ga 6$) planets remain
in terrestrial planet regions, because the damping leads to
isolation between the planets.
We have carried out a series of N-body simulations including
the random torques with different disk surface density
and strength of turbulence.
We found that the orbital 
eccentricities pumped up by the turbulent torques and associated 
random walks in semimajor axes tend to delay isolation of planets,
resulting in more coagulation of planets.
The eccentricities are still damped after planets become isolated. 
As a result, the number of final planets decreases with
increase in strength of the turbulence, while
Earth-mass planets with small eccentricities are still formed.
In the case of relatively strong turbulence,
the number of final planets are 4--5 at 0.5--2AU, which is
more consistent with Solar system,    
for relatively wide range of disk surface density
($\sim 10^{-4}$--$10^{-2}$ times MMSN).
 
\vspace{1em}
keywords: planetary formation, planet-disk interactions, terrestrial planets

\clearpage

\section{Introduction}

The final stage of terrestrial planet accretion would be
coagulation among protoplanets \citep[e.g.,][]{lis87}.
The protoplanets form through oligarchic growth \citep{kok98,kok00},
so that they are called 'oligarchs.'
The protoplanets have almost circular orbits initially and
are isolated from one another \citep{kok98,kok00}.
Their mass is about Mars mass \citep{kok98}
in the case of the minimum mass solar nebula (MMSN) model \citep{hay81}.
Long term distant perturbations, however, would pump up
eccentricities large enough for orbit crossing,
on timescales that depend on mass of protoplanets
and their orbital separation \citep{cha96}.
Because of relatively strong eccentricity damping due to
tidal interaction with a gas disk
\citep{art93,war93}, the orbit crossing may not
occur until disk gas surface density $\Sigma_{\rm g}$ decreases below 
$10^{-3} \Sigma_{\rm g,MMSN}$ \citep{iwa02} where $\Sigma_{\rm g,MMSN}$
is the surface density of MMSN.

N-body simulations without any effect of disk gas 
\citep[e.g.,][]{cha98,agn99} show that 
Earth-mass terrestrial planets are formed at $\sim 1$AU in
a disk with a solid surface density $\sim \Sigma_{\rm d,MMSN}$
as a result of the
orbit crossing but with too large orbital eccentricities ($\sim 0.1$)
to be consistent with the present eccentricities of Earth and Venus
in our Solar system. \citet{kom02,kom04} performed N-body
simulations, taking into account 
the eccentricity damping caused by tidal interaction with a remnant gas disk
and found that final eccentricities can be small enough to
be consistent with those of
Earth and Venus.  The remnant disk with $\Sigma_{\rm g} = 10^{-4}$--
$10^{-3} \Sigma_{\rm g,MMSN}$ allows orbit crossing, but
it is still enough to damp eccentricities of Earth-mass planets
to $\la 0.01$ within disk depletion timescale 
$\sim 10^6$--$10^7$ years \citep{kom02,agn02}.

However, \citet{kom02,kom04} found that generally
$\ga 6$ planets remain in terrestrial planet region
in strong damping cases.
If the damping is weaker, number of planets decreases, while
resultant eccentricities increase.
Only in a limited range of the parameters,
it sometimes occurs that 
Earth-mass planets with small enough eccentricities 
(Earth-like planets) are formed and total number of formed planets
is $\la 4$--5 in a disk similar to MMSN.

\citet{cha01}, \citet{obr06}, and \citet{ray06}
neglected the effects of a gas disk in their N-body simulations, but
included dynamical friction from remnant planetesimals.  
Although the effect of the
dynamical friction is essentially the same as the damping due to
a gas disk, the number of formed planets is fewer
while they have relatively small eccentricities.
More detailed calculations are needed to clarify the role of
dynamical friction from planetesimals.

Another possibility to reduce the number of formed planets 
while their eccentricities are kept small
is a "shaking-up" process to inhibit isolation of planets due to
the damping.  
If gas giants have already formed when the orbital crossing
starts, eccentricity excitation by
the secular perturbations from the gas giants
can provide the shakes \citep[e.g.,][]{lev03}. 
\citet{kom04} showed that perturbations from
gas giants in the current orbits do not provide enough shakes
in the presence of disk gas.
However, \citet{obr06} reported that the
secular perturbations from Jupiter and Saturn
produce terrestrial planets more consistent with 
those in our Solar system in the case without the gas damping
but with dynamical friction from planetesimals.

\citet{nag05} considered a passage of a secular resonance
during depletion of disk gas as a shaking-up mechanism and carried out
N-body simulations.  
$\nu_5$ resonance passes
through $\sim 1$AU when $\Sigma_{\rm g} \sim 10^{-3}$--$10^{-2}\Sigma_{\rm g,MMSN}$.
Hence, after the eccentricity excitation 
and coagulation of protoplanets caused by the
resonance passage, the disk gas is still 
able to damp the eccentricities (Eq.~\ref{eq:t_damp}).
The damping induces inward orbital migration, so that
bodies are captured by the resonances.
The merged terrestrial planets, however, have to be
released from the resonances at $\sim 1$AU.

Here, we consider another shaking-up mechanism, random torques exerted by
disk turbulence due to Magneto-Rotational-Instability (MRI)
\citep[e.g.,][]{bal91}.
\citet[hereafter referred to as LSA04]{lau04}
and \citet{nel04}
carried out fluid dynamical simulations of MRI and
pointed out that the random torques may significantly 
influence orbital motions of planetesimals.
\citet{ric03} studied
accretion of a protoplanet taking into account the effect of random walks
in semimajor axes induced by the random torques 
and found that the random walks expand effective
feeding zone of the protoplanet and it may lead to
rapid formation of a large core of a gas giant.  
However, since they did not integrate orbits directly, it is not 
clear if their incorporation of random torques is relevant.  
Actually, \citet[hereafter referred to as N05]{nel05}
directly integrated orbits of
protoplanets in a turbulent disk and found excitation of orbital
eccentricities as well as random walks of semimajor axes.
The eccentricity excitation was neglected in \citet{ric03}.
Because N05's orbital integration was done simultaneously with 
fluid dynamical simulation of MRI, 
the orbital integration was limited to 100--150 Keplerian times,
which is too short to study accretion process of the
protoplanets on $\ga 10^6$ Keplerian times.

In order to perform N-body simulations
long enough to calculate full stage of accretion of
protoplanets, we adopt the semi-analytical formula 
for the random torque developed by LSA04
based on their fluid dynamical simulations.  
We directly incorporate the random torques as forces
acting on the protoplanets in the equations of motion.  
Hence, eccentricity excitation, which was neglected in
\citet{ric03}, is automatically included,
as well as a random walk in semimajor axis.   
Because we do not perform fluid dynamical simulation, 
N-body simulations on timescales
$\sim 10^7$ years are able to be done.
The analytical formula may only roughly mimic the
effects of MRI turbulence and the method by N05
is more correct.
However, our purpose is rather to explore the qualitative effects
of the turbulence on orbital evolution and accretion of planets 
on long timescales and 
the dependence on the key parameters on the problem, so that
a great quantitative accuracy is not important in the
present contribution.

We also include the damping of eccentricities and
inclinations directly in orbital integrations as forces
acting on the protoplanets, 
essentially following \citet{kom02,kom04}.
However, we here adopt more exact forms of forces derived
by \citet{tan04} (also see \citet{kom05} and section 2.4).
Since in the turbulence, the effect of type-I migration might be
greatly diminished (N05), 
we performed both simulations with type-I migration and
those without it.
When type-I migration is included, we adopt
the formula for forces acting on the protoplanets derived 
by \citet{tan04}.

In section 2, we describe the disk model, the formula of forces
for the random torques, eccentricity damping and type-I migration.
In section 3.1, we present the results of one planet case in order to 
clearly see the effect of the random torques that we use,  
on the orbital evolution.
The results of N-body simulations of accretion of
protoplanets that start with 15 protoplanets of $0.2M_{\oplus}$
are shown in section 3.2.
We mainly consider the stage in which disk gas surface density
has declined significantly so that the random torques are 
relatively weak.
However, the weak random torques in the stage play important
roles to produce terrestrial planets similar to those 
in our Solar system.
Section 4 is the conclusion section.

\section{Model and Calculation Methods}

\subsection{Disk model}

Here, we consider a host star with $1M_{\odot}$.
Following \citet{ida04},
we scale the gas surface density $\Sigma_{\rm g}$ of disks as
\begin{eqnarray}
\Sigma_{\rm g} = 2400 f_g (\frac{r}{1\mathrm{AU}})^{-3/2} \,\mathrm{g\, cm}^{-2},
\label{eq:Sigma}
\end{eqnarray}
where $f_g$ is a scaling factor; 
$f_g=1$ corresponds to $\Sigma_{\rm g} = 1.4 \Sigma_{\rm g,MMSN}$. 
Because current observations cannot strictly constrain the radial
gradient of $\Sigma_{\rm g}$, we here assume $f_g$ is constant with $r$.
Since we consider the stage where disk gas has been
significantly depleted ($f_g \le 10^{-2}$), optical depth
of the disk may be low.
For simplicity, we use the temperature distribution 
in the limit of an optically thin disk \citep{hay81},
\begin{eqnarray}
T = 2.8 \times 10^2 (\frac{r}{1\mathrm{AU}})^{-1/2} \, \mathrm{K}.
\end{eqnarray}
Corresponding sound velocity is 
\begin{eqnarray}
c_s = 1.0 \times 10^5 (\frac{r}{1\mathrm{AU}})^{-1/4} \, \mathrm{cm \, s^{-1}}.
\label{eq:sound_vel}
\end{eqnarray}

\subsection{Random Torques due to MRI Turbulence}

Turbulence due to Magneto-Rotational Instability \citep{bal91} 
is one of candidates to account for the observationally 
inferred disk viscosity, based on H$\alpha$ line observation
due to disk accretion onto stars \citep[e.g.,][]{har98}. 
\citet{gammie96} and \citet{san00} pointed out the existence of "dead zone" 
around 1AU where the degree of ionization is so small that MRI turbulence 
is suppressed.  However, it is not clear that the dead zone exists 
in the last stage of terrestrial planet formation we are considering.
If large fraction of dust grains have been transferred into
planetesimals in the stage, ionization degree could be high enough 
that the dead zone disappears \citep{san00}, 
although secondary dust production due to disruptive
collisions between planetesimals may also be efficient in this stage
\citep[e.g.,][]{inaba03}.
On the other hand, \citet{inutsuka_sano05} proposed 
self-sustained ionization to suggest that dead zone vanishes,
irrespective of degree of dust depletion.
We here assume that disks are MRI-turbulent at $\sim 1$AU. 

LSA04 modeled the density fluctuations
due to the MRI turbulence, based on their fluid dynamical simulation.
Here we briefly summarize their results with slight modifications.
The specific force due to density fluctuations 
exerted on a planet is given by
\begin{equation}
\textbf{\textit{F}}_{{\rm tub}} = - \Gamma \nabla \Phi,
\label{eq:F_tub}
\end{equation}
where
\begin{eqnarray}
\Gamma & = & \frac{64 \Sigma_{\rm g} r^2}{\pi^2 M_{\odot}},
\label{eq:cap_gamma} \\
\Phi & = & \gamma r^2 \Omega^2 \sum_{i=1}^{50} \Lambda_{c,m}, 
\label{eq:phi_MRI} \\
\Lambda_{c,m} & = & 
\xi e^{-\frac{(r-r_{c})^2}{\sigma ^2}} 
\cos (m \theta -\phi_c - \Omega _{c} \tilde{t}) 
\sin (\pi \frac{\tilde{t}}{\Delta t}).
\label{eq:Lambda_c}
\end{eqnarray}
In the above,  
$m$ is the wavenumber, the dimensionless variable $\xi$ 
has a Gaussian distribution with unit width, 
$r$ and $\theta$ represent the location of the planet
in cylindrical coordinates,
$\Omega = \sqrt{GM_{\odot}/r^3}$ is Keplerian angular velocity at $r$, 
$r_c$ and $\phi_c$ specify the center of the density fluctuation,
and $\Omega_{c}$ is $\Omega$ at $r_c$.
$\gamma$ is the non-dimensional 
parameter to indicate the strength of the turbulence
that we introduced instead of LSA04's dimensional parameter $A$
($A = \gamma r^{5/2}\Omega^2$).
The pattern speed $\Omega _c$ in the time-dependent factor 
allows the mode center to travel along with the Keplerian flow.
With $m$ specified, the mode extends for a distance $2 \pi r_c/m$ 
along the azimuthal direction. The radial extent is then specified 
by choosing $\sigma = \pi r_c/4 m$ so that the mode shapes have roughly 
a $4:1$ aspect ratio.
The total fluctuation is expressed by 
superposition of 50 modes at any given time.  
In Eq.(\ref{eq:phi_MRI}), $i$ in $\sum_{i=1}^{50}$ expresses
individual modes.
Each mode comes in and out of existence with 
the time dependence specified above.
An individual mode begins at time $t_0$ and fades away 
when $\Delta t > \tilde{t} \equiv t - t_0$. 
The duration of the mode $\Delta t$ is taken to be the sound crossing 
time of the mode along the angular direction, i.e., 
$\Delta t = 2 \pi r_c/m c_s$. 
After the mode has gone, a new mode is generated.   
For the new mode, $r_c$ is chosen randomly in the
calculation area and $\phi$ is random in $0 \leq \phi < 2\pi$.
The azimuthal wavenumber $m$ is chosen to be distributed according to 
a log random distribution for wavenumbers in the range $2 \leq m \leq 64$.

\begin{figure}
\begin{center}
\plotone{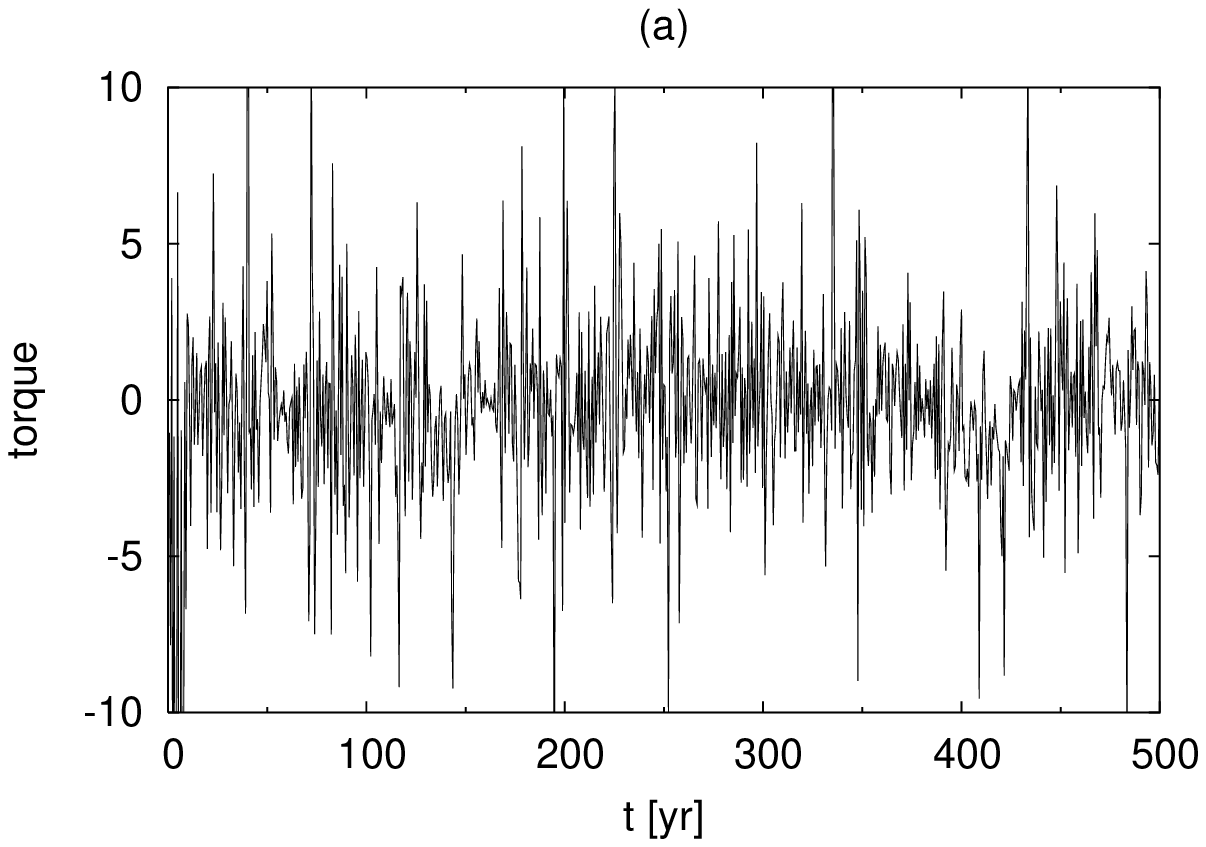}
\plotone{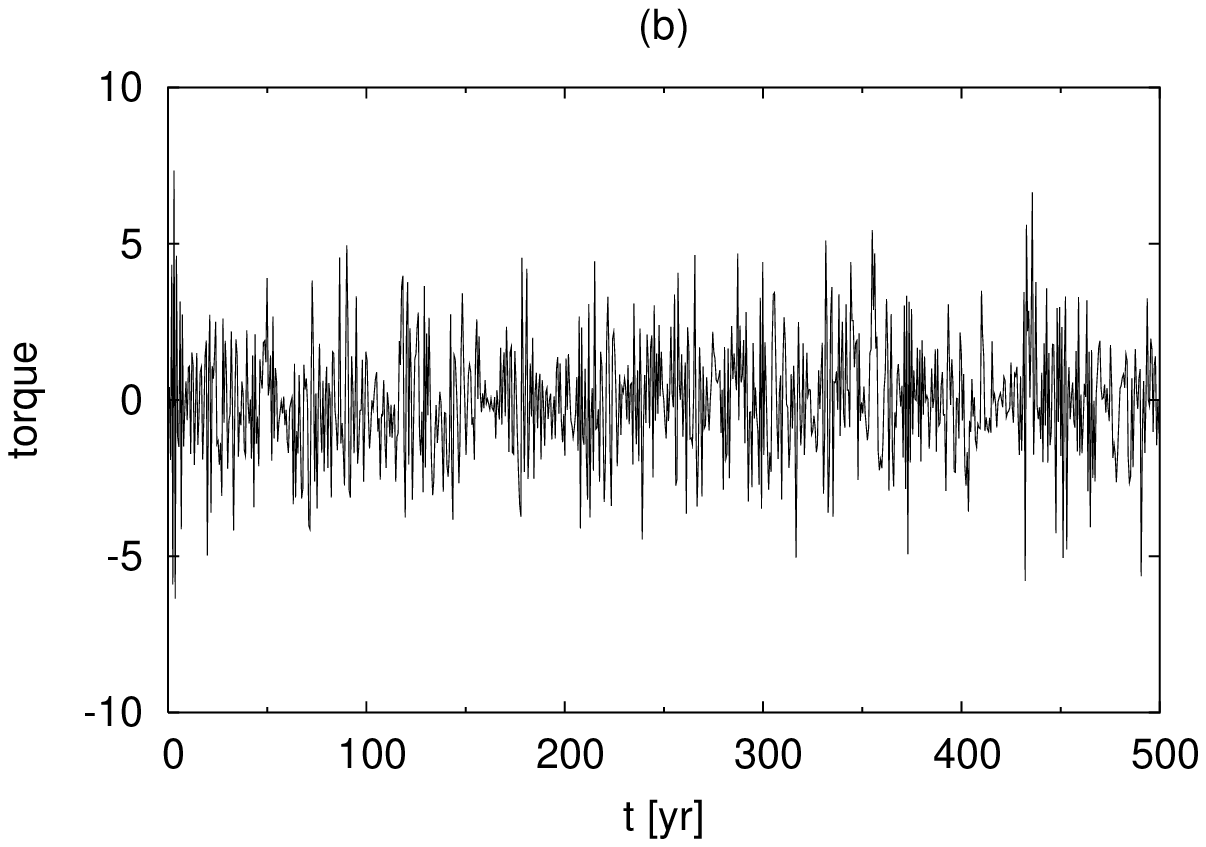}
\caption{
An example of the random torques 
that we adopted.
The (scaled) total torque $\sum_1^{50} m \Lambda_{s,m}$ is plotted
(see Eq.~\ref{eq:tau_tub}),
as a function of time $t$:
(a) the case of $2 \leq m \leq 64$ and 
(b) the case with the modes of $m \geq 6$ omitted (see section 2.4).
}
\label{fig:random_torque}
\end{center}
\end{figure}

We modify their formula in two points.
The first one is introduction of 
the non-dimensional parameter $\gamma$ for strength of turbulence.
While LSA04's simulation range was 1.5--3.5 AU,
our simulations are done around 1AU.
Hence, it may be more useful to use
the non-dimensional parameter than LSA04's dimensional parameter $A$
($A = \gamma r^{5/2}\Omega^2$).
Since LSA04 used 3.4 AU and 1 year as units of length and time 
and the middle radius of their simulation was 2.5 AU,
the values of $A$ in their paper correspond to 
$\simeq 1.2 \gamma (r/1{\rm AU})^{-1/2}$.
Three dimensional fluid dynamical simulations 
by LSA04 suggest $\gamma \sim 10^{-3}$--$10^{-2}$, but
the values of $\gamma$ may include large uncertainty, 
so that we explore wide
range of $\gamma$ (also see discussion in section 3.1).
The second point is the range of wavenumber $m$.
Although $2 \leq m \leq 64$ in the original formula given in LSA04, 
inclusion of $m = 1$ modes could be more consistent with
global fluid dynamical simulation (G. Laughlin, private communication;
also see discussion in section 3.1). 
On the other hand, modes with large $m$ do not contribute to
orbital changes, because of their 
short distances $(\sim \sigma = \pi r_c/4 m)$ for effective forces 
and rapid time variations with timescale $2 \pi r_c/m c_s$.
Hence, in order to save calculation time, we cut off 
high $m$ modes with $m \geq 6$, that is, 
$\Lambda_{c,m}$ is set to be zero when $m \geq 6$,
in the summation of 50 modes in Eq.~(\ref{eq:phi_MRI}) 
The torque exerted on the planet with mass $M$ is
\begin{eqnarray}
\tau_{{\rm tub}} &=& r \times \frac{1}{r} \frac{\partial \Phi}{\partial \theta} \times \Gamma \times M \\ 
&=& - \gamma \Gamma M r^2 \Omega ^2 \sum_{i=1}^{50} m \Lambda_{s,m}, 
\label{eq:tau_tub} 
\end{eqnarray}
where
\begin{equation}
\Lambda_{s,m} = \xi e^{-\frac{(r-r_{c})^2}{\sigma ^2}} 
\sin (m \theta - \phi_c - \Omega _{c} \tilde{t}) \sin (\pi \frac{\tilde{t}}{\Delta t}). 
\label{eq:Lambda_s}
\end{equation}
Figure \ref{fig:random_torque}a shows 
the scaled random torques, $\sum_{i=1}^{50} m \Lambda_{s,m}$, with
$2 \leq m \leq 64$.
In Fig.~\ref{fig:random_torque}b, $m \geq 6$ modes are cutted off.
Low frequency patterns, which contribute to
orbital evolution, are similar between these results.
Actually, 
we found through orbital calculations like in Fig.~\ref{fig:one_body_random}
that $m \leq 5$ are enough to reproduce orbital evolution
with fully counting all $m$ modes.
If $m = 1$ mode is included, 
the amplitude of the random torques does not change, but
low frequency patterns change.
We will present the results of N-body simulations with 
$2 \leq m \leq 5$ in section 3.2, but
we also carried out calculations with $1 \leq m \leq 5$ and
will discuss the effects of $m = 1$ modes.

\begin{center}
[Figure \ref{fig:random_torque}]
\end{center}

\subsection{Secular Torques due to Disk-Planet Interactions}

As shown below, the random torques given by Eq.~(\ref{eq:tau_tub})
induce random walks of semimajor axes of planets 
and pump up their orbital eccentricities.
Tidal interactions with a laminar disk 
monotonically decreases the semimajor axes and 
damp the eccentricities (and the inclinations).
The secular inward migration is known as "type-I migration"
\citep[e.g.,][]{war86,war97,tan02}.
Since mean flow in turbulent disks coincides with 
flow in laminar disks, 
interactions with the mean flow may induce the secular
orbital migration and eccentricity damping even in turbulent disks.
In turbulent disks, however, N05 reported that
type-I migration might be greatly diminished 
while the eccentricity damping still works.
Non-linear effects associated with the random
fluctuations (e.g., the temporary activation of corotation
torques or temporary disruption of the pressure buffer)
could be responsible for the slowing down. 
Alternatively, relatively high eccentricities
excited by the random torques,
which is $\ga h/r \sim 0.05$ obtained by N05
where $h$ is disk scale height, could affect the type-I migration
\citep[e.g.,][]{Pap00}.
In our N-body simulations, 
eccentricities are also pumped up to $\ga 0.05$ 
by perturbations among protoplanets 
except for the last phases well after orbital crossing.
Hence, we performed two series of simulations:
one is without type-I migration and 
the other is with it. 

We summarize the secular changes in laminar disks below.
Both torques from inner and outer disks damp orbital
eccentricities and inclinations, since the gravitational
interactions with disk gas causes similar effect of
dynamical friction.  The damping timescales are \citep{tan04}
\begin{eqnarray}
t_{{\rm damp}, e} & = & - \frac{e}{\dot{e}} = \frac{t_{\rm damp}}{0.78} \\
t_{{\rm damp}, i} & = & - \frac{i}{\dot{i}} = \frac{t_{\rm damp}}{0.54} \\
t_{\rm damp} & = & (\frac{M}{M_{\odot}})^{-1} 
               (\frac{\Sigma_{\rm g} a^2}{M_{\odot}})^{-1}
               (\frac{c_s}{v_{\rm K}})^4 \Omega^{-1} \\
             & = & 240 f_g^{-1} (\frac{M}{M_{\oplus}})^{-1} 
                (\frac{a}{1{\rm AU}})^2 {\rm years},
\label{eq:t_damp}
\end{eqnarray}
where $a$ is the semimajor axis of the planet and $v_{\rm K}$ is the Keplerian 
velocity at $a$.

On the other hand, the torque from an inner disk increases
semimajor axis, while that from an outer disk decreases it.
Since the outer torque is generally greater than the inner one
\citep{war86,tan02},
the torque imbalance induces inward migration (type-I migration).
For the radial gradient of $\Sigma_{\rm g} \propto a^{-1.5}$,
the torque imbalance, which is negative definite, 
is given by \citep{tan02}
\begin{equation}
\tau_{\rm mig} = -2.17(\frac{M}{M_{\odot}})^2 (\frac{v_{\rm K}}{c_s})^2 \Sigma_{\rm g} a^4 \Omega^2.
\label{eq:torque_mig}
\end{equation}
Migration timescale due to this torque is
\begin{eqnarray}
t_{\rm mig} = - \frac{a}{\dot{a}} 
& = & \frac{(1/2) M \Omega a^2}{\tau_{\rm mig}} \\
& = & 0.23
(\frac{M}{M_{\odot}})^{-1}
(\frac{\Sigma_{\rm g} a^2}{M_{\odot}})^{-1}
(\frac{c_s}{v_{\rm K}})^{2} \Omega^{-1} \\ 
& = & 5.0 \times 10^4 
(\frac{M}{M_{\oplus}})^{-1} (\frac{a}{1 \, \mathrm{AU}})^{3/2} f_g^{-1} \, \mathrm{years}. 
\label{eq:t_mig}
\end{eqnarray}

\subsection{Orbital Integration}

We integrate orbits of 15 protoplanets with $0.2M_{\oplus}$
that initially have orbits of small $e$ and $i$ ($\sim 0.01$) with
separation  $6r_{\rm H}$, following initial conditions
in \citet{kom02}, where
Hill radius $r_H$ is defined by
\begin{equation}
r_H = (\frac{M}{3M_{\odot}})^{1/3} a 
\simeq 0.007(\frac{M}{0.2M_{\oplus}})^{1/3} a.
\end{equation}
Initial angular distributions are set to be random.
Calculation starts from the phase when the orbital crossing starts. 
The result of \citet{kok00} shows that the eccentricities 
of protoplanets produced through oligarchic growth
are about $\sim 10^{-3}$, so that
the protoplanets are well isolated.
However, the protoplanets will eventually start orbital crossing by 
long-term mutual distant perturbations on a time scale depending on 
their orbital separation, mass \citep{cha96}, 
initial eccentricities \citep{yos99}, 
and how much gas is around the protoplanets \citep{iwa02}. 
Since we are concerned with orbital crossing stage,
we start the calculation with relatively high eccentricities
$e = 10^{-2}$, supposing the eccentricities have already increased 
and orbital crossing is ready to start. 
The initial inclinations are also set to be $i=10^{-2}$.

The basic equations of motion of particle $k$ at $\textbf{\textit{r}}_k$ 
in heliocentric coordinates are
\begin{eqnarray}
\frac{d^2 \textbf{\textit{r}}_k}{dt^2} 
& = & -GM_{\odot} \frac{\textbf{\textit{r}}_k}{ |\textbf{\textit{r}}_k|^3} 
- \sum_{j \neq k} GM_j 
\frac{\textbf{\textit{r}}_k - \textbf{\textit{r}}_j}{|\textbf{\textit{r}}_k - \textbf{\textit{r}}_j|^3} 
- \sum_{j} GM_j \frac{\textbf{\textit{r}}_j}{|\textbf{\textit{r}}_j|^3} 
\nonumber\\
  &   &   
+ \textbf{\textit{F}}_{\rm damp} + \textbf{\textit{F}}_{\rm tub} + \textbf{\textit{F}}_{\rm mig},
\label{eq:2} 
\end{eqnarray}
where $k,j = 1,2,...,15$, the first term is gravitational force 
of the central star, 
the second term is mutual gravity between the bodies, 
and the third term is the indirect term.
$\textbf{\textit{F}}_{\rm damp}$ and $\textbf{\textit{F}}_{\rm mig}$ 
are specific forces for the damping of eccentricities and inclinations
and type-I migration,
and $\textbf{\textit{F}}_{{\rm tub}}$ is specific force 
due to the turbulence (Eq.~\ref{eq:F_tub}).
Their detailed expressions are described below.
Note that in our simulations, mass of bodies is larger than $0.2M_{\oplus}$,
so that aerodynamical drag forces are neglected compared with 
$\textbf{\textit{F}}_{\rm damp}$ and $\textbf{\textit{F}}_{\rm mig}$
\citep[e.g.,][]{war93}. 

We integrate orbits with the fourth-order Hermite scheme. 
When protoplanets collide, perfect accretion is assumed. 
After the collision, a new body is created,
conserving total mass and momentum of
the two colliding protoplanets.
The physical radius of a protoplanet is determined by its mass and internal density as
\begin{equation}
r_P = (\frac{3}{4 \pi} \frac{M}{\rho_P})^{1/3}.
\end{equation}
The internal density $\rho_P$ is set to be $3 \,\mathrm{g\, cm}^{-3}$.

\citet{tan04} derived, through three-dimensional linear analysis,
\begin{eqnarray}
\textit{F}_{{\rm damp},r} 
   & = & (\frac{M}{M _{\odot}}) (\frac{v_{\rm K}}{c _s})^4 
    (\frac{\Sigma_{\rm g} r^2}{M _{\odot}}) 
    \Omega (2A ^{c}_{r}[v _{\theta} - r \Omega] + A ^{s}_{r} v _r) \label{eq:3-1} \\
\textit{F}_{{\rm damp}, \theta} 
   & = & (\frac{M}{M _{\odot}}) (\frac{v_{\rm K}}{c _s})^4 
   (\frac{\Sigma_{\rm g} r^2}{M _{\odot}})
   \Omega (2A ^{c}_{\theta}[v _{\theta} - r \Omega] + A ^{s}_{\theta} v _r) \label{eq:3-2}  \\
\textit{F}_{{\rm damp},z} 
   & = & (\frac{M}{M _{\odot}}) (\frac{v_{\rm K}}{c _s})^4 
   (\frac{\Sigma_{\rm g} r^2}{M _{\odot}}) \Omega (A^{c}_{z} v_z + A ^{s}_{z} z \Omega ) \label{eq:3-3} \\
\textit{F}_{{\rm mig},r} & = & 0 \label{eq:3-1-2} \\
\textit{F}_{{\rm mig}, \theta} 
   & = &  - 2.17(\frac{M}{M_{\odot}}) (\frac{v_{\rm K}}{c _s})^2 
   (\frac{\Sigma_{\rm g} r^2}{M _{\odot}}) \Omega v_{\rm K} 
   \label{eq:3-2-2} \\
\textit{F}_{{\rm mig},z} & = & 0, \label{eq:3-3-2}
\end{eqnarray}
where 
\begin{eqnarray}
A^{c}_{r} = 0.057  & & A^{s}_{r} = 0.176 \nonumber \\
A^{c}_{\theta} = -0.868  & & A^{s}_{\theta} = 0.325 \nonumber \\
A^{c}_{z} = -1.088  & & A^{s}_{z} = -0.871 .\nonumber  
\end{eqnarray}
Note that there is a typos in $\textit{F}_{{\rm damp},z}$
in \citet{tan04}.
The factor $(2A^{c}_{z} v_z + A ^{s}_{z} z \Omega)$
should be $(A^{c}_{z} v_z + A ^{s}_{z} z \Omega)$ as in
Eq.~(\ref{eq:3-3}).
Note also that the other factors in the 
expressions in \citet{kom05} have
minor typos; the above expressions are correct ones.
Eccentricities are damped by $\textit{F}_{{\rm damp},r}$ and
$\textit{F}_{{\rm damp},\theta}$, while 
inclinations are damped by $\textit{F}_{{\rm damp},z}$.
Semimajor axes are decreased by 
$\textit{F}_{{\rm mig},\theta}$ $(= \tau_{\rm mig}/Mr)$
where $a$ and $r$ are
identified because of small $e$ and $i$.
The evolution of $e, i$ and $a$ by orbital integration 
of one body with the above forces
completely agrees with the analytically derived evolution
with Eqs.~(\ref{eq:t_damp}) and (\ref{eq:t_mig}).

The force due to turbulence, 
$\textbf{\textit{F}}_{{\rm tub}} = - \Gamma \nabla \Phi$
(Eq.~\ref{eq:F_tub}), is given by
\begin{eqnarray}
\textit{F}_{{\rm tub}, r} & = & 
  \gamma \Gamma r \Omega^2 
  \sum_{i=1}^{50} \left( 1 + \frac{2r(r-r_c)}{\sigma^2}\right)
  \Lambda_{c,m}, \label{eq:f_tub_r} \\
\textit{F}_{{\rm tub}, \theta} & = & 
  \gamma \Gamma r \Omega^2 \sum_{i=1}^{50} m \Lambda_{s,m},
  \label{eq:f_tub_theta} \\
\textit{F}_{{\rm tub}, z} & = & 0,
\end{eqnarray}
where $\Lambda_{c,m}$ and $\Lambda_{s,m}$ are 
defined by Eqs.~(\ref{eq:Lambda_c}) and (\ref{eq:Lambda_s}). 

As seen above,
$\textbf{\textit{F}}_{\rm damp}$ and $\textbf{\textit{F}}_{\rm mig}$
are parameterized by the disk surface scaling factor $f_g$ for 
given $M$ and $r$ of planets,
and $\textbf{\textit{F}}_{{\rm tub}}$ by $f_g$ and $\gamma$
($\Gamma \propto f_g$).
Therefore, $f_g$ and $\gamma$ are parameters for our calculations.
As discussed in section 1, we will consider the stages in which
disk gas has been significantly depleted, so that 
the cases of $f_g = 10^{-4}$ and $10^{-2}$ 
are mainly studied.
Although the most likely value of $\gamma$ might be $\sim 10^{-3}$--$10^{-2}$,
it would include large uncertainty
(also see discussion in section 3.1), so that
the cases of $\gamma = 10^{-3}, 10^{-1}$ and 1 are studied.  
For comparison, non-turbulent ($\gamma = 0$) cases are also calculated.

\section{Results}

\subsection{One Planet Case}
\label{subsec:one_planet}

\begin{figure}
\begin{center}
\plotone{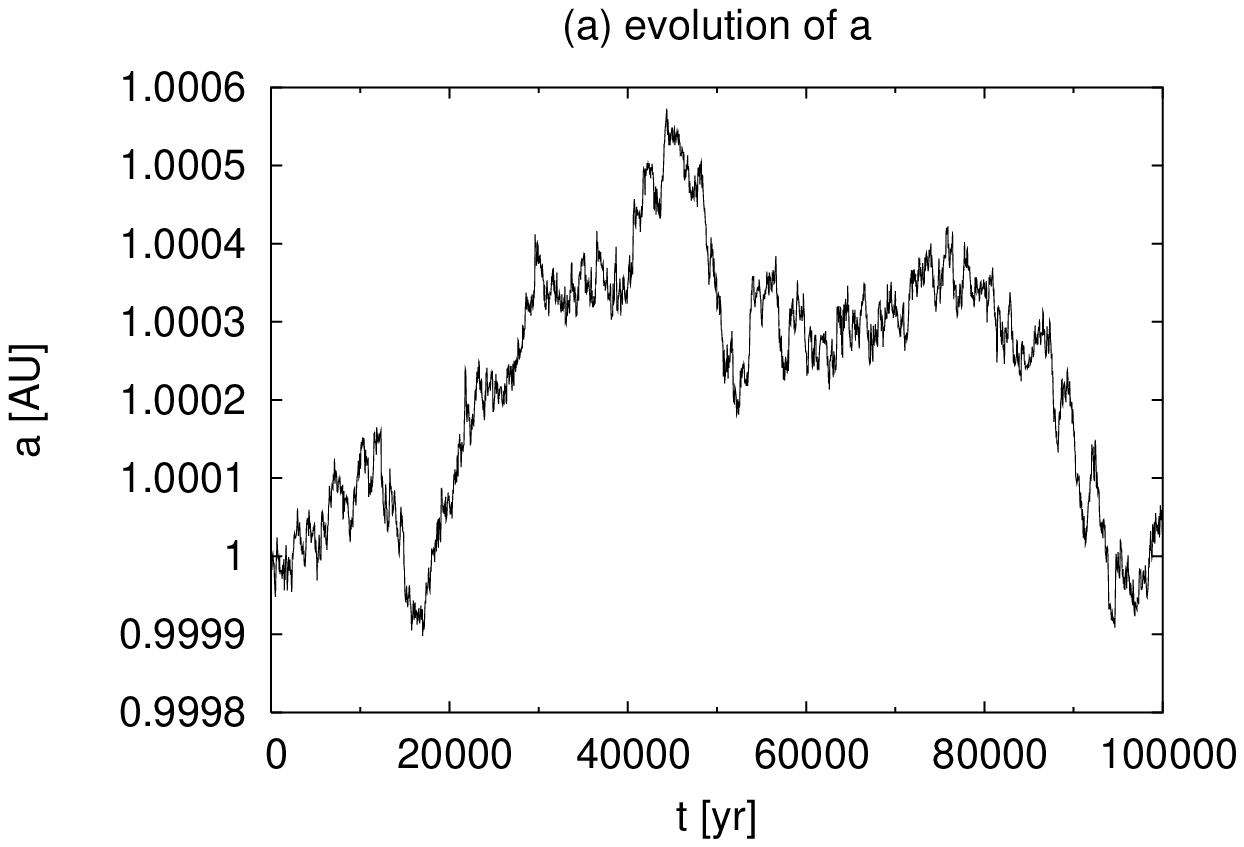}
\plotone{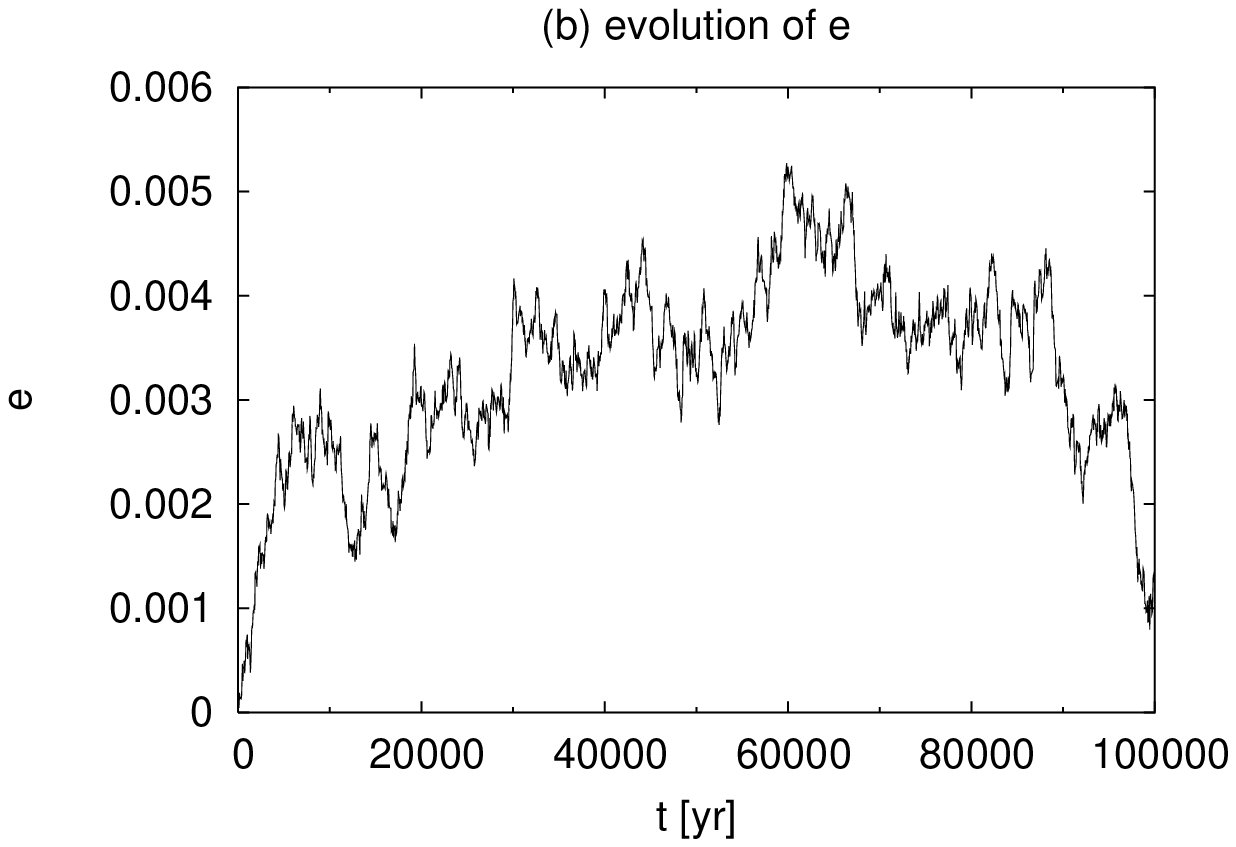}
\caption{
Orbital evolution of a planet of $0.2M_{\oplus}$ 
suffering random torques due to turbulent fluctuations
with $f_g = 10^{-2}$ and $\gamma = 10^{-1}$.
Type-I migration is not included.
(a) evolution of semimajor axis $a$ and (b) that of orbital eccentricity $e$.
}
\label{fig:one_body_random}
\end{center}
\end{figure}

To see the effects of the turbulent forces
given by Eqs.~(\ref{eq:f_tub_r}) and (\ref{eq:f_tub_theta}) on
orbital changes and how they depend on $f_g$ and $\gamma$, 
we first carry out simulations with one planet embedded in
a turbulent disk.
Figures~\ref{fig:one_body_random} show 
evolution of semimajor axis $a$ and orbital eccentricity $e$ 
of a planet of $0.2M_{\oplus}$ obtained by
orbital integration with 
$\textbf{\textit{F}}_{{\rm tub}}$ 
in the case of $\gamma = 10^{-1}$ and $f_g = 10^{-2}$. 
The initial conditions are $a=1$AU and $e=0$.
$\textbf{\textit{F}}_{\rm damp}$ and 
$\textbf{\textit{F}}_{\rm mig}$ are not included. 
As expected, a random walk of $a$ and 
excitation of $e$ are observed.

\begin{center}
[Figure \ref{fig:one_body_random}] 
\end{center}

\begin{figure}[htbp]
\begin{center}
\plotone{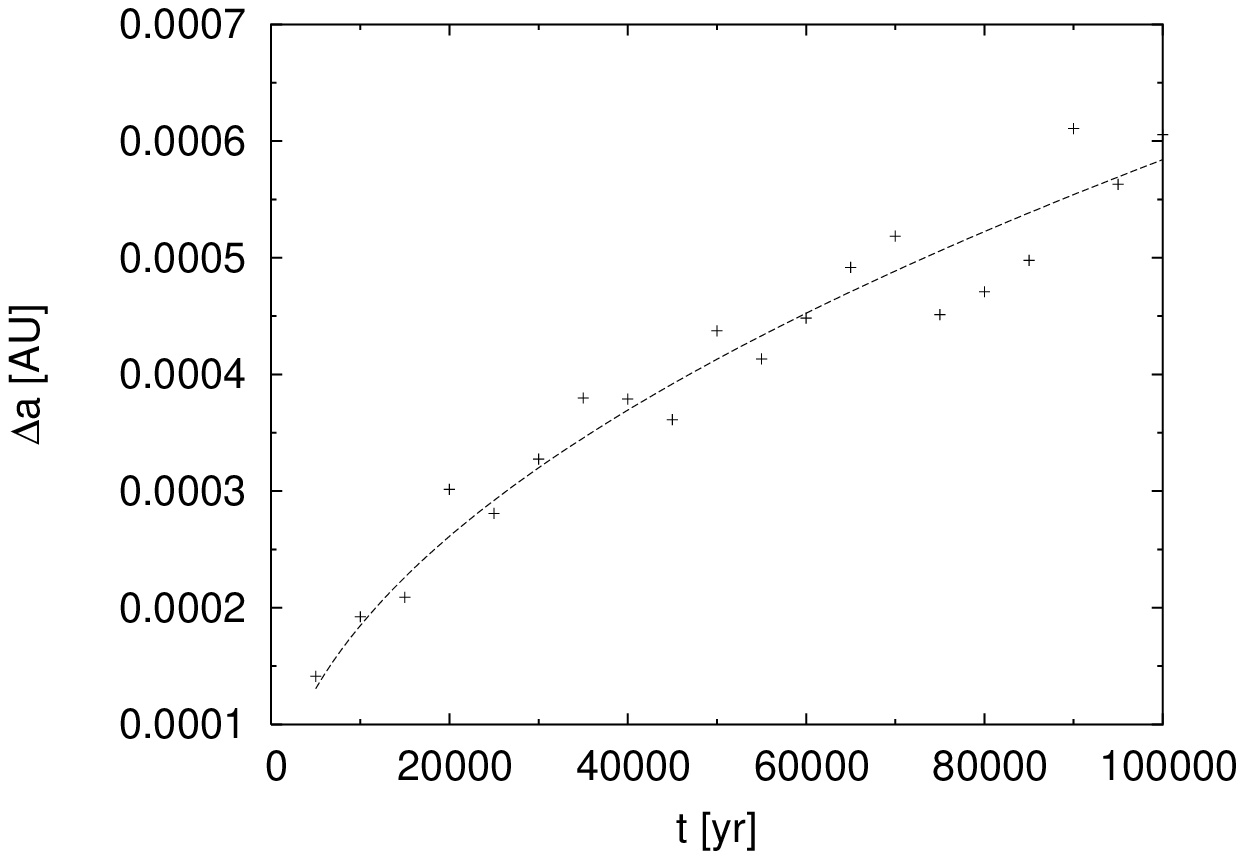}
\plotone{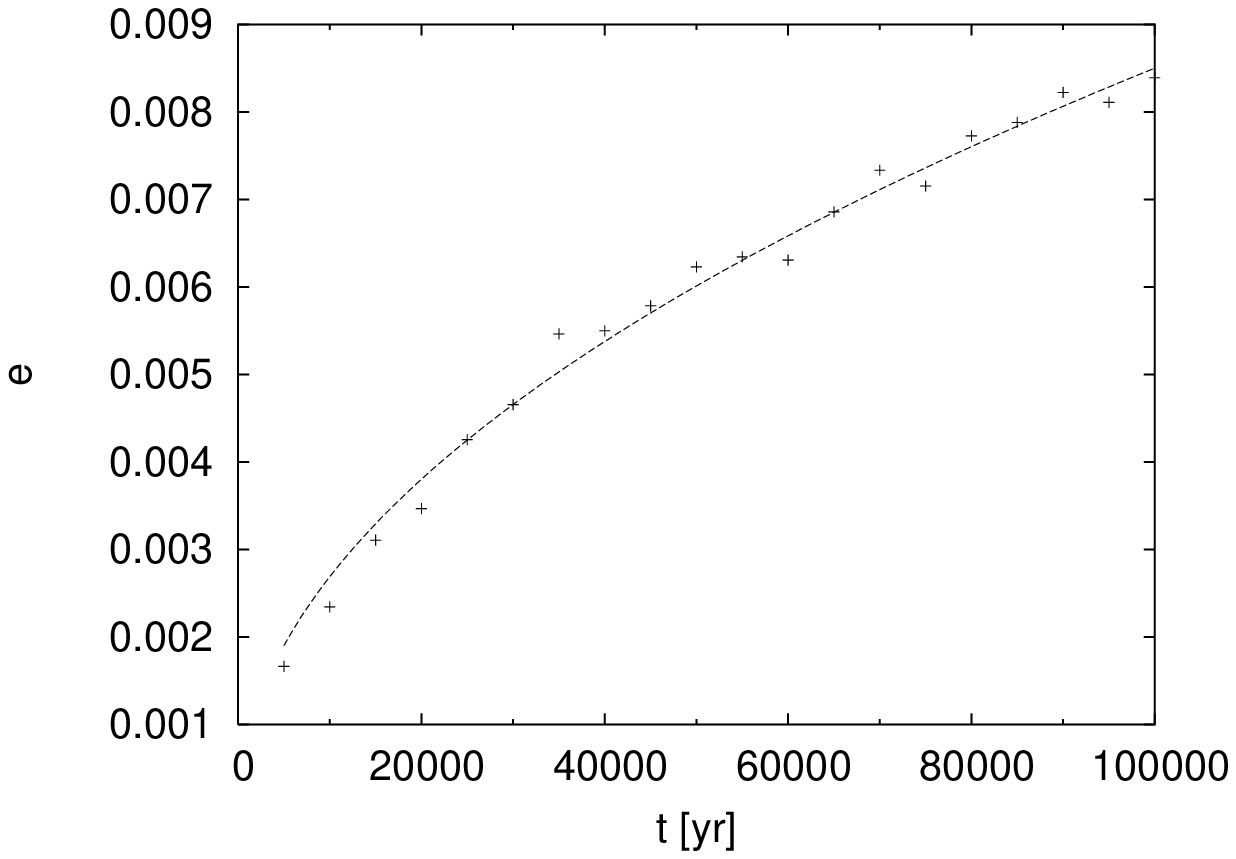}
\caption{
Time evolution of dispersion of $\Delta a$ and $e$.
Crosses are the standard deviations of 
Gaussian distributions derived from 100 numerical integrations
with different random numbers for random torques 
but with $f_g = 10^{-2}$ and $\gamma = 10^{-1}$.
The solid curves are fitted functions given by
Eqs.~(\ref{eq:random_walk_fitting}) and 
(\ref{eq:e_excite_fitting}).
}
\label{fig:2}
\end{center}
\end{figure}

\begin{center}
[Figure \ref{fig:2}] 
\end{center}

In order to quantify the random walks,
we performed 100 similar runs with different random numbers
for the random torques, but still using 
$\gamma = 10^{-1}$ and $f_g = 10^{-2}$. 
At each time, the distributions of deviation in semimajor axis 
$\Delta a$ from the initial position (1AU) and orbital eccentricity $e$
for the 100 runs are
fitted as Gaussian distributions 
to obtain the standard deviations as functions of time.
Hereafter, the standard deviations are also denoted by
$\Delta a$ and $e$.
Figures~\ref{fig:2} show the evolution of $\Delta a$ and $e$
obtained by the numerical calculations.
The evolution curves are fitted as
\begin{eqnarray}
\Delta a & \sim & 1.8 \times 10^{-6} (\frac{t}{1{\rm year}})^{1/2}{\rm AU},
\label{eq:random_walk_fitting} \\
e & \sim & 2.7 \times 10^{-5} (\frac{t}{1{\rm year}})^{1/2}.
\label{eq:e_excite_fitting} 
\end{eqnarray}
near 1AU.
The dependence of $t^{1/2}$ would reflect diffusion characteristics.
(If $\textbf{\textit{F}}_{\rm damp}$ is included, $e$ approaches
an equilibrium value.)
We have carried out the same procedures
for $\gamma = 10^{-2},10^{-1}$ and
$f_g = 10^{-2},10^{-1}$ to derive the dependence of $\gamma$ and $f_g$ as
\begin{eqnarray}
\Delta a & \sim & 2 \times 10^{-3} f_g \gamma 
  (\frac{t}{1{\rm year}})^{1/2}{\rm AU}.
\label{eq:random_walk_t_dep} \\
e & \sim & 3 \times 10^{-2} f_g \gamma 
  (\frac{t}{1{\rm year}})^{1/2}.
\label{eq:e_excite_t_dep} 
\end{eqnarray}
Note that the random walks are independent of planet mass $M$.
In the above calculations, we used $m = 2$--5 modes.
With $m = 2$--64, we obtained very simular results.

Equations (\ref{eq:random_walk_t_dep}) and
(\ref{eq:e_excite_t_dep}) give
$\Delta a$ and $e$ that are 10--100 times smaller 
than those obtained by LSA04 and N05.
We found that the analytically modeled random torques almost 
cancel out in time and the net change is only $\sim 0.001$
of total change for $m = 2$--5.  
Since both LSA04 and N05 used 
global fluid codes to follow orbits of protoplanets, 
$m = 1$ modes might be included.
Since $m = 1$ modes have the longest duration and 
distance for effective force, it would
induce asymmetry between positive and negative
torques to produce larger $\Delta a$ and $e$.
We have carried out similar calculations, including 
$m = 1$ modes and found that
$\Delta a$ and $e$ are 10 times larger than
Eqs.~(\ref{eq:random_walk_t_dep}) and (\ref{eq:e_excite_t_dep}).
The inclusion of $m = 1$ modes may mostly resolve
the difference from the results by LSA04 and N05, but
the approximated semi-analytical torque formula could 
be still too symmetric,
compared with the global fluid dynamical simulations.  
Hence, the results of $\gamma = 10^{-1}$ and 1 (with $m = 2$--5 modes),
which are larger than the numerically inferred 
value $\gamma \sim 10^{-3}$--$10^{-2}$,
are also pertinent for the evolution of planets
in realistic turbulent disks.
In the N-body simulations shown in section 3.2,
perturbations from other protoplanets also induce some asymmetry
and $\Delta a$ and $e$ may be much larger than
Eqs.~(\ref{eq:random_walk_t_dep}) and (\ref{eq:e_excite_t_dep})
during the period in which protoplanets
undergo relatively close encounters.

\begin{figure}
\begin{center}
\plotone{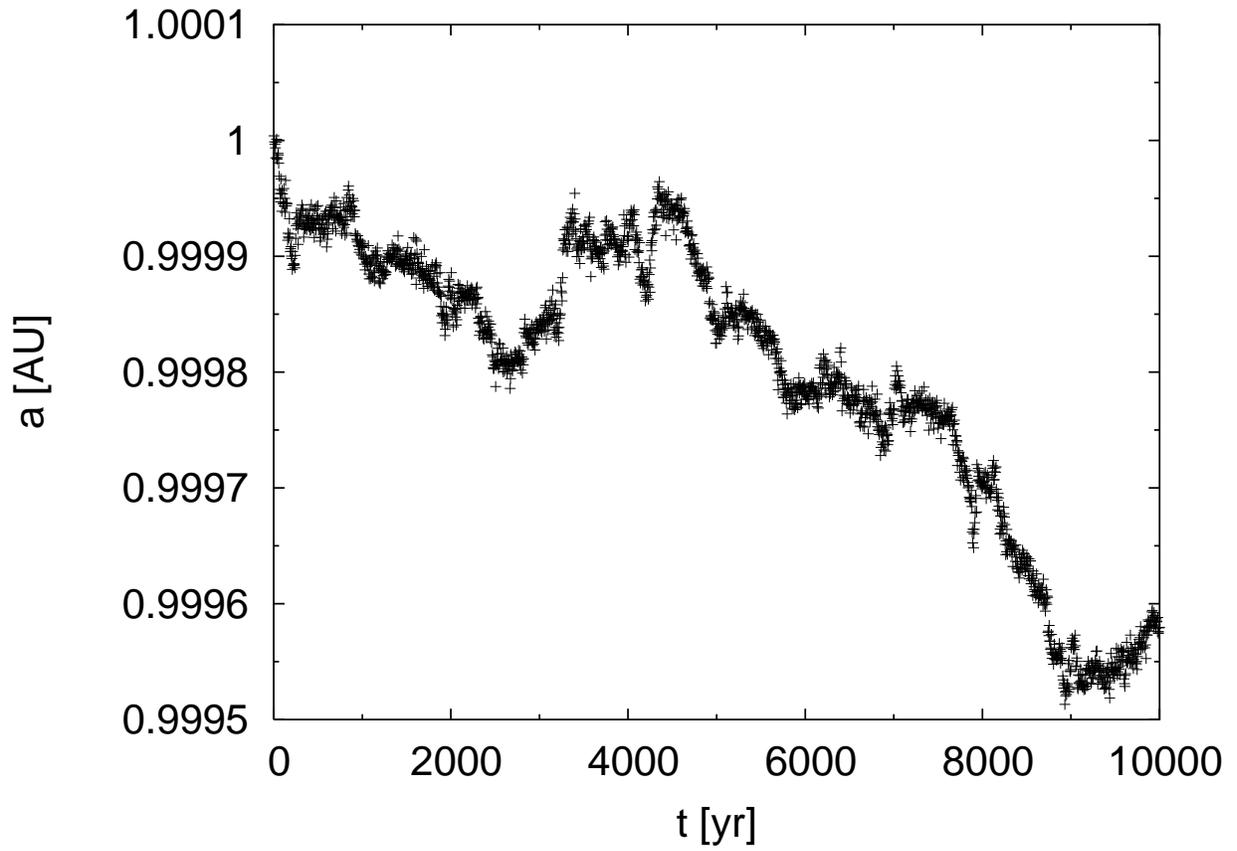}
\caption{
Orbital evolution with 
both effects of type-I migration and turbulent fluctuations
with $f_g = 10^{-2}$ and $\gamma = 10^{-1}$.
}
\label{fig:4}
\end{center}
\end{figure}

If type-I migration works on the timescale given by 
Eq.~(\ref{eq:t_mig}), the migration length near 1AU is
\begin{equation}
\Delta a \sim 2 \times 10^{-5} f_g (\frac{M}{M_{\oplus}})
           (\frac{t}{1{\rm year}}) {\rm AU}.
\label{eq:type_I_t_dep} 
\end{equation}
From Eqs.~(\ref{eq:random_walk_t_dep}) and (\ref{eq:type_I_t_dep}), 
it is expected that if
\begin{eqnarray}
t \ga 3 \times 10^5 \gamma^2 (\frac{M}{0.2M_{\oplus}})^{-2} \, \mathrm{years} ,
\label{eq:critical_t_for_type_I}
\end{eqnarray}
type-I migration will dominate over the random walk.
Figure \ref{fig:4} shows the evolution of the semimajor axis 
with both effects of type-I migration and turbulent fluctuations
of $\gamma = 0.1$.
The planet starts secular inward migration after $t \sim 3 \times 10^3$ years,
consistent with the above estimate.
Note, however, that in the turbulent disks, it is not clear that
type-I migration speed is still the same as that predicted by
the linear calculation (N05).

\begin{center}
[Figure \ref{fig:4}] 
\end{center}

\subsection{Accretion of Protoplanets in a Turbulent Disk}

Because we will compare the results with \citet{kom02}
and because type-I migration might be greatly diminished 
in turbulent disks,
in many runs we calculate accretion and the orbital evolution of protoplanets
in a turbulent disk without the effect of type-I migration.
We carry out simulations with various $f_g$ and $\gamma$. 
We denote a run with $f_g = 10^{-\alpha}, \gamma = 10^{-\beta}$ 
as RUN$\alpha_{\beta k}$, where $k$ ($k = a,b,c$) represent
different initial angular distribution of the protoplanets.
In some runs, the effect of type-I migration is included,
which we denote as RUN $\alpha_{\beta a I}$.
Table 1 shows simulation parameters for individual runs
with $\gamma \leq 1$ and $m = 2$--5 (28 runs).
Two runs were carried out with $\gamma = 10$ ($m = 2$--5).
We also carried out 18 runs with inclusion of $m = 1$ modes
and found that slightly smaller $\gamma$ 
produce similar results to the cases without $m = 1$ modes.
To avoid confusion, we will only present 
the detailed results with $m = 2$--5.

\subsubsection{Case with $f_g = 10^{-2}$}

First we show the results with $f_g = 10^{-2}$.
The orbital evolution of RUN$2_{\infty a}$, RUN$2_{3 a}$, RUN$2_{1 a}$, 
and RUN$2_{0 a}$ are 
shown in Figs.~\ref{fig:fg-2}a, b, c, and d, respectively. 
The thick solid lines represent semimajor axes $a$. 
The thin dashed lines represent pericenters $a(1-e)$ and apocenters $a(1+e)$. 
Thicker solid lines represents more massive planets.
With $f_g = 10^{-2}$, the
damping time scale $\tau _{\rm damp} \simeq 1.2 \times 10^5$ years 
for $M = 0.2M_{\oplus}$. 

\begin{center}
[Figure \ref{fig:fg-2}] 
\end{center}

Since RUN$2_{\infty a}$ does not include the random torques ($\gamma = 0$),
the evolution in Fig.~\ref{fig:fg-2}a is very similar to
that shown by \citet{kom02}.
In this case, 
a planet of $0.6 M_{\oplus}$ with small eccentricity 
($\sim 0.0001$) is formed.
However, global orbital crossing lasts for only $\sim 5 \times 10^5$ years 
because of the rather strong eccentricity damping. 
Consequently, the number of surviving planets are 8, which is
much greater than that in the present Solar system. 
The runs with very weak turbulence of $\gamma = 10^{-3}$
in Fig.~\ref{fig:fg-2}b
shows a similar result to the non-turbulent case.
For $\gamma = 0$ and $10^{-3}$, 
the number of surviving planets is always 8 or 9 (Table 1).

The effects of turbulence are pronounced in the cases of
$\gamma = 10^{-1}$ (Fig.~\ref{fig:fg-2}c)
 and $\gamma = 1$ (Fig.~\ref{fig:fg-2}d).
The random walk and eccentricity excitation induced by the turbulence 
tend to inhibit isolation of the planets.
In Fig.~\ref{fig:fg-2}c (RUN$2_{1 a}$), 
the duration of orbit crossing is elongated,
while 8 planets still survive.
(The same number of planets survive also in RUN$2_{1 b}$ and RUN$2_{1 c}$.)  
According to the eccentricity excitation effect,
the eccentricities of final planets are slightly larger than 
in the previous two cases, however, they are still smaller
than the present free eccentricities of Earth and Venus, because
the damping that increases with planet mass
eventually overwhelms the turbulent excitation that is independent of
the planet mass.
In Fig.~\ref{fig:fg-2}d (RUN$2_{0 a}$ with $\gamma = 1$), 
the large random walk 
enhances the number of collision events (10 events), so that
the number of surviving planets drastically decreases to 4. 
In RUN$2_{0 b}$ and RUN$2_{0 c}$, 
the number of surviving planets is also 4 or 5.

In Fig.~\ref{fig:fg-2}d ($\gamma = 1$), secular inward migration is found,
although type-I migration is not included.
This migration is induced by 
the damping of eccentricities that are continuously
excited by the random torques, since 
orbital angular momentum, $\sqrt{G M_{\odot} a(1-e^2)}$,
is almost conserved during the eccentricities damping.
In the run with extremely large $\gamma$ ($= 10$),
the turbulent excitation is so strong that
all the planets are removed from terrestrial planet region
by the inward migration.

\subsubsection{Case with $f_g = 10^{-4}$}

The evolution in severely depleted disks with $f_g = 10^{-4}$,
RUN$4_{\infty a}$, RUN$4_{3 a}$, RUN$4_{1 a}$, and RUN$4_{0a}$, are
shown in 
Figs.~\ref{fig:fg-4}a, b, c, and d, respectively. 
$f_g = 10^{-4}$ corresponds to $\tau _{\rm damp} \simeq 1.2 \times 10^7$ years
for $M = 0.2M_{\oplus}$.
RUN$4_{\infty a}$ shows the result without the effect of turbulence.
The weak damping due to the small surface density of a gas disk 
elongates the period during which the eccentricities are high enough to
allow orbital crossing ($\sim 1 \times 10^7$ years). 
As a result, a larger planet ($M = 1.4 M_{\oplus}$)
than in RUN$2_{\infty a}$ is formed.
Since $e$ is damped down to $\sim 0.01$, this planet is 
very similar to Earth. 
However, the number of surviving planets is 6 (Fig.~\ref{fig:fg-4}a), 
which is larger than that in the present Solar system, as is the case
shown by \citet{kom02}. 
The mean number of surviving planets of 
RUN$4_{\infty a}$, RUN$4_{\infty b}$ and RUN$4_{\infty c}$ is 6.3.

\begin{center}
[Figure \ref{fig:fg-4}]
\end{center}

In the turbulence cases,
the mean number of surviving planets is
5.8 ($\gamma = 10^{-3}$), 5.7 ($\gamma = 10^{-1}$), and
4.7 ($\gamma = 1$).
As the turbulence becomes stronger,
the number of final planets decreases.
In RUN$4_{1 a}$ with $\gamma = 10^{-1}$, 
global orbital crossing lasts on more than $10^7$ years (Fig.~\ref{fig:fg-4}c),
while RUN$4_{3 a}$ does not show such clear elongation of orbital
crossing (Fig.~\ref{fig:fg-4}b).
The turbulent excitation for eccentricities is 
still weaker than the tidal damping for Earth-mass planets 
as long as $\gamma \le 1$, so that their final eccentricities
are still $\la 0.01$. (In the run with extremely large $\gamma$  
($= 10$), we found that the eccentricities are not sufficiently damped.)
In general, probability for final planets to be similar to
present terrestrial planets in our Solar system is larger for
$f_g = 10^{-4}$ than for $f_g = 10^{-2}$.

The accretion timescales in weak turbulence cases are 
a few times $10^6$ years after orbit crossing starts.
Those in strong turbulence cases are $\sim 10^7$ years.
\citet{iwa02} and \citet{kom02} suggested
that orbit crossing does not start until $f_g$ decays 
down to $\sim 10^{-3}$.
If the effect of turbulence is taken into account,
orbit crossing may start at the stage of larger $f_g$.
If the condition of $f_g \la 10^{-3}$ is applied and exponential decay 
from initial $f_g \sim 1$ with
decay timescale $\tau_{\rm dep}$ is assumed,
the orbit crossing starts 
at $\sim 7\tau_{\rm dep} \sim 10^7$--$10^8$ years.
Thus, the total accretion timescales in the present model
are not in contradiction to the Earth formation age
inferred from Hf-W chronology $\sim 4 \times 10^7$ years
\citep{yin02, yin03, kleine02, kleine04}. 

If disk depletion is only due to viscous diffusion,
it may be possible that such small-mass remnant disks remain
for $10^7$--$10^8$ years.
However, if disk dispersal due to stellar EUV
is efficient, it would be difficult to preserve
such small-mass remnant disks.
Spitzer survey found that 25\% of B-A members
and 10\% of F-K members in Pleiades cluster show IR excess
\citep{nadya06}.
The excess might imply the existence of small-mass remnant 
gas disks, but it might also be due to 
secondary dust generation in gas free environments.
Observation of gas components for clusters at
$10^7$--$10^8$ years is needed to examine
the role of the tidal damping due to remnant disks
on final orbital configuration of terrestrial planets.

\subsubsection{Eccentricities and Feeding Zones}

\begin{figure}[htbp]
\begin{center}
\plotone{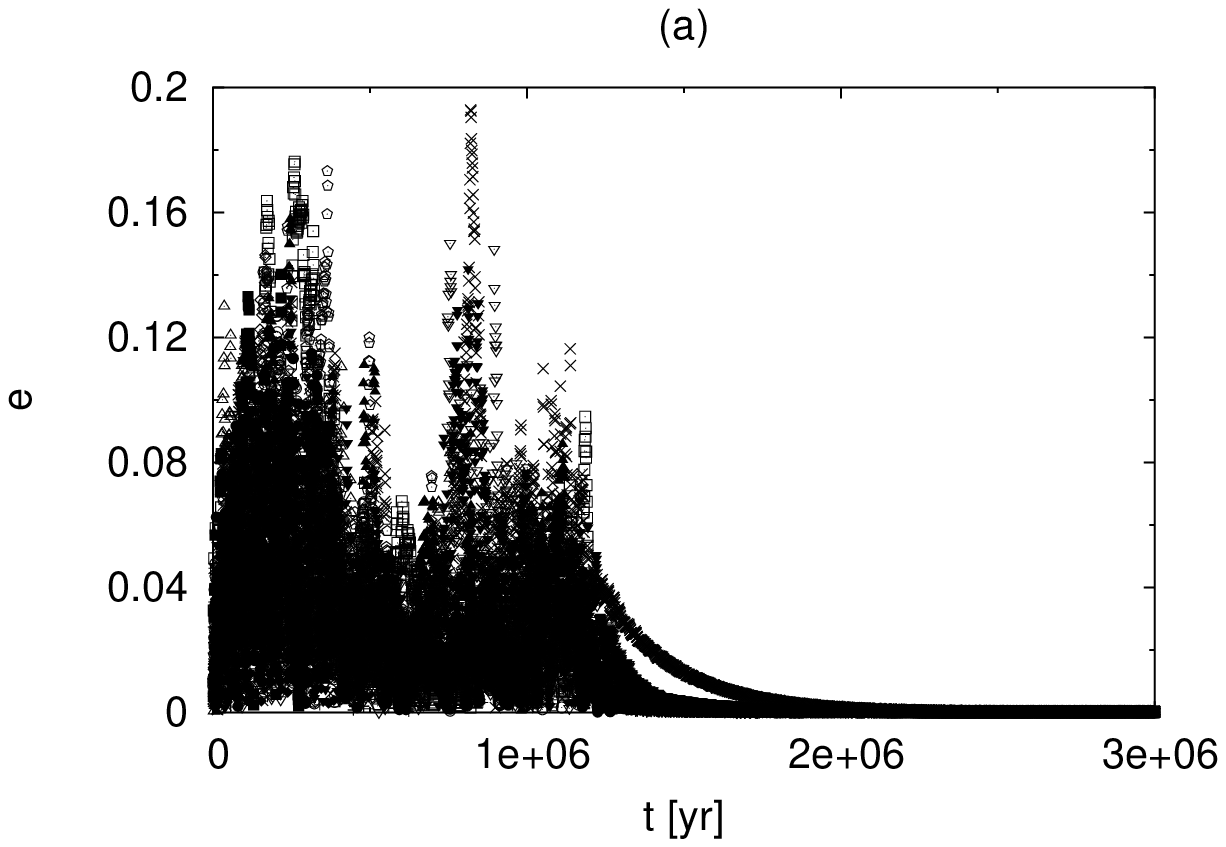}
\plotone{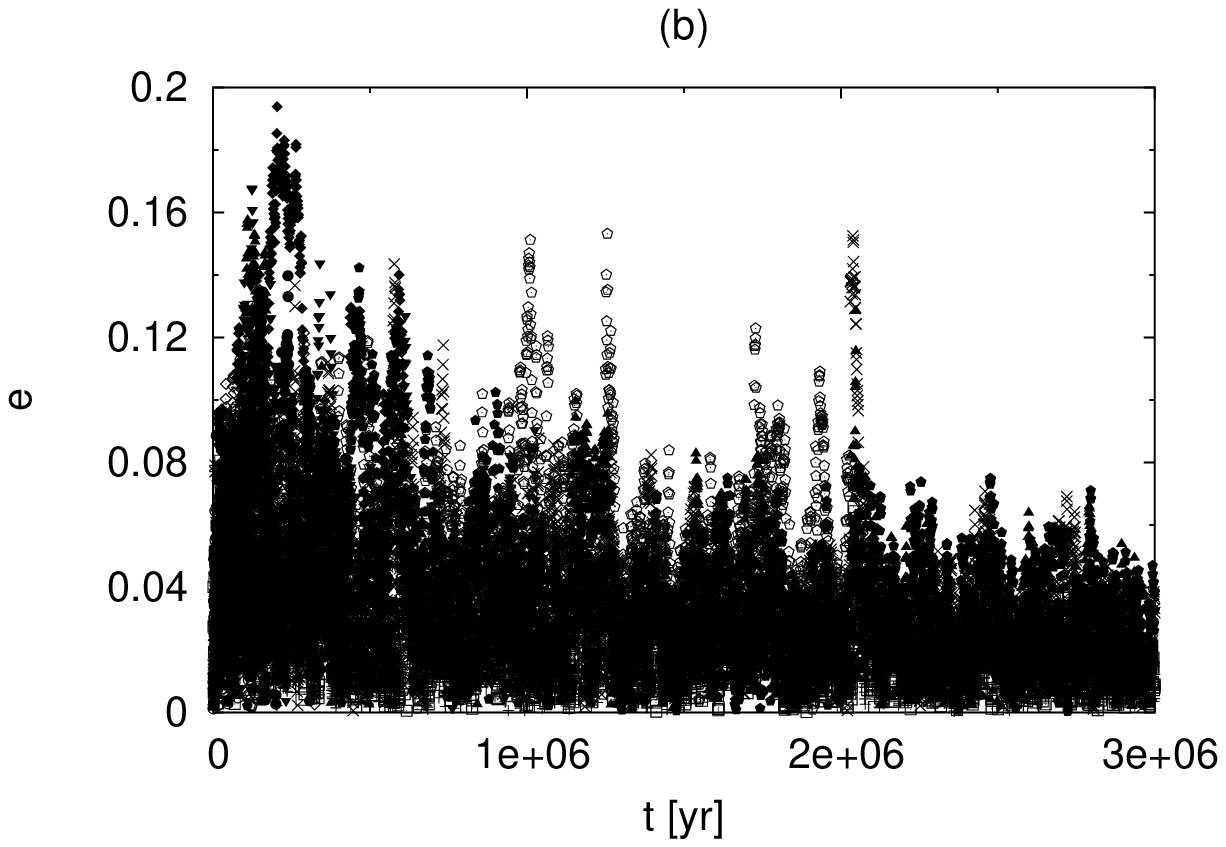}
\caption{
The eccentricity evolution of all bodies: 
(a) $\gamma = 0$ [RUN$2_{\infty a}$] and 
(b) $\gamma = 1$ [RUN$2_{0a}$].
}
\label{fig:4-13}
\end{center}
\end{figure}

Once orbit crossing starts, the velocity dispersion 
is pumped up to surface escape velocity $v_{\rm esc}$ of planets
by close encounters.  Corresponding eccentricity is given by
\begin{equation}
e \sim \frac{v_{\rm esc}}{v_{\rm K}} 
  = 0.34 (\frac{\rho_P}{3{\rm gcm}^{-3}})^{1/6}
    (\frac{M}{M_{\oplus}})^{1/3} (\frac{a}{1{\rm AU}})^{1/2}.
\label{eq:e_v_esc}
\end{equation}
Figure \ref{fig:4-13} shows  
eccentricity evolution of all bodies in RUN$2_{\infty a}$ 
(non-turbulent case) and RUN$2_{0a}$ (strongly turbulent case
with $\gamma = 1$).
During orbit crossing ($t \la 1 \times 10^6$ years),
the mean eccentricities are almost same in the two cases.
Since eccentricity excitation due to random torques evaluated by
Eq.~(\ref{eq:e_excite_t_dep}) is significantly smaller than Eq.~(\ref{eq:e_v_esc}), 
the eccentricities during orbit crossing are mostly determined by
mutual planetary perturbations.

\begin{center}
[Figure \ref{fig:4-13}]
\end{center}

In RUN$2_{\infty a}$, 
global orbit crossing ceases at $t \ga 1 \times 10^6$ years
and then the eccentricities are secularly decreased by
the damping due to $\textbf{\textit{F}}_{\rm damp}$. 
However, in RUN$2_{0a}$, close encounters still occasionally
occur at $t \ga 1 \times 10^6$ years, so that 
the eccentricity damping is slower.
Even with relatively strong turbulence of $\gamma=1$ of this run,
the tidal damping of eccentricities
overcomes the turbulent excitation for Earth-mass planets.
(But, this is not the case for $\gamma=10$.)

For $\gamma=1$ and $f_g = 10^{-2}$,
diffusion length due to random torques is evaluated by
Eq.~(\ref{eq:random_walk_t_dep}) as
$\Delta a \sim 10^{-2} (t/10^{6}{\rm year})^{1/2}{\rm AU}$,
which is much smaller than orbital separation among the
protoplanets.
However, as suggested before, planetary perturbations
may inhibit cancellation of the torques and induce
much larger $\Delta a$ (and $e$).
Furthermore, even if the turbulence itself does not 
directly expand the feeding zones of the planets,   
scattering by close encounters among protoplanets 
that are induced by 
the random torques allows the feeding zones to effectively expand.
In the N-body simulations, the two effects are
indistinguishable.

\subsubsection{Effects of Type I Migration}

Here, the results 
with type-I migration (calculations with
$\textbf{\textit{F}}_{\rm mig}$) are shown, although
it is not clear that type-I migration actually operates
in turbulent disks (N05).
In RUN$2_{\infty aI}$, $f_g = 10^{-2}$ and
the turbulence is not included ($\gamma = 0$).
More systematic investigations in the non-turbulence cases were
done by \citet{mcn05} and \citet{kom06}.
Although \citet{mcn05} and \citet{kom06} included
the effects of small planetesimals as well, RUN$2_{\infty aI}$ shows
similar properties to their calculations
(Fig.~\ref{fig:fg-2I}a): planets
in inner regions tend to fall onto the host stars while
those in outer regions could survive.
Since collision events are limited by the loss of inner planets,
the mass of the largest surviving planets is $0.4M_{\oplus}$. 
(Inclusion of protoplanets in more outer region might increase
the mass of final planets.)

\begin{center}
[Figure \ref{fig:fg-2I}]
\end{center}

Figure~\ref{fig:fg-2I}b and \ref{fig:fg-2I}c show 
RUN$2_{1 aI}$ of $\gamma = 10^{-1}$ and 
RUN$2_{0 aI}$ of $\gamma = 1$. 
Even with relatively strong turbulence, 
the tendency to migrate inward does not change,
compared with the non-turbulent case in Fig.~\ref{fig:fg-2I}a.
Equation~(\ref{eq:critical_t_for_type_I}) shows that 
type-I migration is dominant 
over the random walk after 
$t \sim 3 \times 10^5 \gamma^2 (M/0.2M_{\oplus})^{-2}$ years, so that
the random walk cannot halt the inward migration on timescales
$\sim 10^6$ years.
The inward migration is rather accelerated by 
the damping of eccentricities that are continuously
excited by the random torques (section 3.2.1).

For larger $f_g$, since the random torques are stronger,
the accelerated migration is more pronounced.
Thus, our results suggest that random
migration superposed to type-I migration would not be able to solve
the problem that planets tend to be lost from the terrestrial planet
region. The problem can be solved only if the turbulent fluctuations
somehow inhibit the underlying type-I migration, as found by N05,
or if planets at $\sim 1$AU are formed by surviving protoplanets
originally at $> 1$ AU.

\section{Conclusions}

We have investigated the final accretion stage of terrestrial planets 
from Mars-mass protoplanets in turbulent disks, through N-body simulation.
Gravitational interactions with gas disks
exert the following three effects on the protoplanet orbits:
\begin{enumerate}
\item damping of eccentricities $e$ and inclinations $i$,
\item type-I migration (secular decrease of semimajor axis $a$),
\item random-walks of $a$ and stochastic excitation of $e$.
\end{enumerate}
The effect 3) has not been included in N-body simulations
of planet accretion in the previous works.
We adopt the same simulation setting of \citet{kom02}
that included only the effect 1): initially 15 protoplanets of 
$0.2M_{\oplus}$ are set with orbital separations of several Hill
radii in terrestrial planet regions, corresponding to MMSN.  
In our N-body simulations, the effects 1) and 3) 
were included.
The effect 2) was examined in section 3.2.4.
We incorporated random torques exerted by
disk turbulence due to MRI as forces directly acting on protoplanets
in the equations of motion for orbital integration.
We adopted the semi-analytical formula for
the random torques developed by LSA04 with slight modifications.  
Compared with the results of \citet{kom02},
we investigated the effects of disk turbulence on planet accretion.  

The past N-body simulations neglecting the gas disk
showed that the coagulation between protoplanets result in planets of 
about Earth-mass but with the eccentricities higher than
the present terrestrial planets in our Solar system.
If the effect 1) is included, 
when $\Sigma_{\rm g} \sim 10^{-4}$--$10^{-3} \Sigma_{\rm g,MMSN}$,
the damping allows initiation of orbit crossing
to form an Earth-mass planet(s),
while it damps the eccentricities sufficiently after planets
are isolated. 
However, $\ga 6$ planets tend to remain, 
because of isolation due to the damping \citep{kom02}.
(Note that in the case of damping by dynamical friction from
remnant planetesimals the number of planets is reduced \citep{obr06}).
We found that the newly incorporated effect 3)
tends to inhibit isolation of planets,
resulting in more coagulations of planets, while the eccentricity
damping is still effective.
As a result, 4--5 planets
with small eccentricities are formed in 
relatively wide parameter range: 
gas surface density 
$\Sigma_{\rm g} \sim 10^{-4}$--$10^{-2} \Sigma_{\rm g,MMSN}$,
and MRI turbulence strength $\gamma \sim 10^{-1}$--1
(slightly smaller $\gamma$ 
if $m = 1$ modes of density fluctuation are included).

LSA04's prescription for the random torques 
that we adopted has highly symmetric properties
and the exerted torques almost completely cancel out
in time averaging.
As a result, the diffusion length and eccentricity excitation 
obtained by one planet calculations are
generally too small to play a direct role in
expanding feeding zones of protoplanets
in late phase in which disk gas is significantly depleted
(Eqs.~\ref{eq:random_walk_t_dep} and \ref{eq:e_excite_t_dep}).
However, planetary perturbations may break the symmetry.
Furthermore, in more realistic turbulence, the torques may 
include $m = 1$ modes and be less symmetric.  These effects
inhibit the torque cancellation
to induce much larger $\Delta a$ and $e$.
Furthermore, even if the enhanced effects are still too small
to directly expand the feeding zones,
such small effects can be enough to break the isolation of the
protoplanets, thus allowing them to have distant encounters with each
other. The encounters in turn induce larger random oscillations of the
semimajor axes, effectively enhancing the feeding zone of each planet.
We also found through calculations with all the effects 1), 2) and
3) that the random walks do not decelerate (rather accelerate)
the type-I migration, although
it is not clear that
type-I migration actually operates in turbulent disks.

Although the prescription for the random torques would include
large uncertainty, we have demonstrated that
the random torques tend to decrease number of final planets
while they keep formation of Earth-mass planets with small eccentricities,
which is more consistent with the present Solar system.
Since the random torques are independent of mass of bodies,
small planetesimals also suffer the random torques.
N-body simulations starting from smaller planetesimals
in turbulent disks will be presented in a separate paper.
 
\vspace{1em} 
\noindent 
{\bf Acknowledgments}.  We thank Fred Adams and Greg Laughlin for 
detailed and useful comments on the random torque functions.
We also thank an anonymous referee for helpful comments.

\clearpage

\begin{table}
  \label{tbl:1}
  \begin{center}
    \caption{Initial parameters and final results for each run.
$M_1$ is the mass of the largest planet in final state. $e_1$ is the time averaged eccentricity of the largest planet, taken after its isolation takes place.}

\begin{small}    
    \begin{tabular}{l c c c c c c c} \hline
      RUN & $f_g$ & $\gamma$ & type-I & $ M_1 (M _{\oplus})$ &$e_1$ & collision events & number of final planets \\ \hline
      RUN$2_{\infty a}$ & $10^{-2} $ & 0  & No &0.6&0.0001&7&8\\ 
      RUN$2_{\infty b}$ & $10^{-2} $ & 0  & No &0.8&0.0004&7&8\\ 
      RUN$2_{\infty c}$ & $10^{-2} $ & 0  & No &0.8&0.0001&6&9\\ 
      RUN$2_{3 a}$ & $10^{-2} $ & $ 10^{-3}$  & No &0.6&0.002&7&8\\ 
      RUN$2_{3 b}$ & $10^{-2} $ & $ 10^{-3} $  & No &0.8&0.003&6&9\\ 
      RUN$2_{3 c}$ & $10^{-2} $ & $ 10^{-3} $  & No &0.8&0.0002&6&9\\ 
      RUN$2_{1 a}$ & $10^{-2} $ & $ 10^{-1} $  & No &0.8&0.003&7&8\\ 
      RUN$2_{1 b}$ & $10^{-2} $ & $ 10^{-1} $  & No &0.8&0.003&7&8\\ 
      RUN$2_{1 c}$ & $10^{-2} $ & $ 10^{-1} $  & No &0.8&0.003&7&8\\ 
      RUN$2_{0 a}$ & $10^{-2} $ & $ 1 $  & No &1.2&0.02&10&5\\ 
      RUN$2_{0 b}$ & $10^{-2} $ & $ 1 $  & No &1.0&0.02&11&4\\ 
      RUN$2_{0 c}$ & $10^{-2} $ & $ 1 $  & No &1.0&0.02&10&5\\ 
      RUN$4_{\infty a}$ & $10^{-4} $ &  0   & No &1.4&0.01&9&6\\ 
      RUN$4_{\infty b}$ & $10^{-4} $ &  0   & No &1.2&0.001&7&8\\ 
      RUN$4_{\infty c}$ & $10^{-4} $ &  0   & No &1.2&0.003&10&5\\ 
      RUN$4_{3 a}$ & $10^{-4} $ & $ 10^{-3} $  & No &1.6&0.01&10&5\\ 
      RUN$4_{3 b}$ & $10^{-4} $ & $ 10^{-3} $  & No &1.2&0.01&11&4\\ 
      RUN$4_{3 c}$ & $10^{-4} $ & $ 10^{-3} $  & No &1.2&0.02&8&7\\
      RUN$4_{3 d}$ & $10^{-4} $ & $ 10^{-3} $  & No &1.0&0.02&8&7\\
      RUN$4_{1 a}$ & $10^{-4} $ & $ 10^{-1} $  & No &1.2&0.008&9&6\\ 
      RUN$4_{1 b}$ & $10^{-4} $ & $ 10^{-1} $  & No &1.8&0.004&10&5\\ 
      RUN$4_{1 c}$ & $10^{-4} $ & $ 10^{-1} $  & No &1.0&0.02&9&6\\
      RUN$4_{0 a}$ & $10^{-4} $ & $ 1 $  & No &1.6&0.01&11&4\\ 
      RUN$4_{0 b}$ & $10^{-4} $ & $ 1 $  & No &1.6&0.008&10&5\\ 
      RUN$4_{0 c}$ & $10^{-4} $ & $ 1 $  & No &1.2&0.01&10&5\\ 
      RUN$2_{\infty aI}$ & $10^{-2} $ &  0   & Yes &0.4&0.005&7&3($t=10^7$ years)\\ 
      RUN$2_{3 aI}$ & $10^{-2} $ & $ 10^{-3} $  & Yes &0.2&0.0001&9&1($t=10^7$ years)\\ 
      RUN$2_{1 aI}$ & $10^{-2} $ & $ 10^{-1} $  & Yes &0.4&0.01&6&5($t=10^7$ years)\\ 
      RUN$2 _{0 aI}$ & $10^{-2} $ & $ 1 $  & Yes &0.2&0.02&10&1($t=10^7$ years)\\ 
      \hline
    \end{tabular}
    \end{small}    
  \end{center}
\end{table}

\clearpage

\begin{figure}[htbp]
\begin{center}
\plotone{I09782_fig5.eps3}
\caption{
The results of N-body simulations with $f_g = 10^{-2}$.
(a) $\gamma = 0$ [RUN$2_{\infty a}$], 
(b) $\gamma = 10^{-3}$ [RUN$2_{3 a}$], 
(c) $\gamma = 10^{-1}$ [RUN$2_{1 a}$], 
and (d) $\gamma = 1$ [RUN$2_{0 a}$].
The thick solid lines represent semimajor axes $a$. 
The thin dashed lines represent pericenters $a(1-e)$ and apocenters $a(1+e)$. 
Thicker solid lines represents more massive planets.
}
\label{fig:fg-2}
\end{center}
\end{figure}

\begin{figure}[htbp]
\begin{center}
\plotone{I09782_fig6.eps3}
\caption{
The results of N-body simulations with $f_g = 10^{-4}$.
(a) $\gamma = 0$ [RUN$4_{\infty a}$], 
(b) $\gamma = 10^{-3}$ [RUN$4_{3 a}$], 
(c) $\gamma = 10^{-1}$ [RUN$4_{1 a}$], 
and (d) $\gamma = 1$ [RUN$4_{0 a}$].
The meaning of lines is the same as Figs.~\ref{fig:fg-2}.
}
\label{fig:fg-4}
\end{center}
\end{figure}

\begin{figure}[htbp]
\begin{center}
\plotone{I09782_fig8.eps3}
\caption{
The results of N-body simulations with $f_g = 10^{-2}$,
including type-I migration.
(a) $\gamma = 0$ [RUN$2_{\infty aI}$], 
(b) $\gamma = 10^{-1}$ [RUN$2_{1 aI}$], 
and (c) $\gamma = 1$ [RUN$2_{0 aI}$].
The meaning of lines, see caption of Figs.~\ref{fig:fg-2}.
}
\label{fig:fg-2I}
\end{center}
\end{figure}


\begin{thebibliography}{}

\bibitem[Agnor \& Canup(1999)]{agn99} Agnor, C. B., \& Canup, R. M., 1999, 
On the character and consequences of large impacts in the late stage of terrestrial planet formation,
Icarus, 142, 219

\bibitem[Agnor \& Ward(2002)]{agn02} Agnor, C. B., \& Ward, W. R., 2002,
Damping of terrestrial planet eccentricities by density wave interactions with a remnant gas disk, \apj, 567, 579

\bibitem[Artymowicz(1993)]{art93} Artymowicz, P., 1993,
Disk-satellite interaction via density waves and the eccentricity evolution of bodies embedded in disks, \apj, 419, 166

\bibitem[Balbus \& Hawley(1991)]{bal91} Balbus, S. A., \& Hawley, J. F., 1991,
A powerful local shear instability in weakly magnetized disks. I - Linear analysis. II - Nonlinear evolution, \apj, 376, 214

\bibitem[Chambers et al.(1996)]{cha96} Chambers, J. E., Wetherill, G. W., \& Boss, A. P. 1996,
The stability of multi-planet systems, Icarus, 119, 261

\bibitem[Chambers \& Wetherill(1998)]{cha98} Chambers, J. E., \& Wetherill, G. W., 1998
Making the terrestrial planets: N-body integrations of planetary embryos in three dimensions, Icarus, 136, 304

\bibitem[Chambers(2001)]{cha01} Chambers, J. E., 2001,
Making more terrestrial planets, Icarus, 152, 205

\bibitem[Gammie(1996)]{gammie96} 
Gammie, C. F., 1996,
Layered accretion in T Tauri disks, \apj, 457, 355

\bibitem[Hartmann et al.(1998)]{har98} 
Hartmann, L., Calvet, N., Gullbring, E. \& D'Alessio, P. 1998.
Accretion and the evolution of T Tauri disks, \apj, 495, 385

\bibitem[Hayashi(1981)]{hay81} Hayashi, C., 1981,
Structure of the solar nebula, growth and decay of magnetic fields and effects of magnetic and turbulent viscosities on the nebula, Prog. Theor. Phys. Suppl, 70, 35

\bibitem[Ida \& Lin(2004)]{ida04} Ida, S., \& Lin, D. N. C., 2004,
Toward a deterministic model of planetary formation. I. A desert in the mass and semimajor axis distributions of extrasolar planets, \apj, 604, 388

\bibitem[Inaba et al.(2003)]{inaba03} 
Inaba, S., Wetherill, G. W. \& Ikoma, M., 2003.
Formation of gas giant planets: core accretion models with 
fragmentation and planetary envelope, Icarus, 166, 46-62.

\bibitem[Inutsuka \& Sano(2005)]{inutsuka_sano05} 
Inutsuka, S. \& Sano, T., 2005.
Self-sustained ionization and vanishing dead zones in protoplanetary disks,
ApJL, 628, 155-158.

\bibitem[Iwasaki et al.(2002)]{iwa02} Iwasaki, K., Emori, H., Nakazawa, K., \& Tanaka, H., 2002,
Orbital stability of a protoplanet system under a drag force proportional to the random velocity, Publ. Astron. Soc. Japan, 54 471

\bibitem[Kleine et al.(2002)]{kleine02} Kleine, T., M{\"u}nker,
C., Mezger, K., Palme, H.\ 2002, \ Rapid accretion and early core formation
on asteroids and the terrestrial planets from Hf-W chronometry.\ Nature
418, 952-955.

\bibitem[Kleine et al.(2004)]{kleine04} 
Kleine, T., Mezger, K.,
Palme, H., M{\"u}nker, C.\ 2004.\ The W isotope evolution of the bulk
silicate Earth: constraints on the timing and mechanisms of core formation
and accretion.\ Earth and Planetary Science Letters 228, 109-123.

\bibitem[Kokubo \& Ida(1998)]{kok98} Kokubo, E., \& Ida, S. 1998,
Oligarchic growth of protoplanets, Icarus, 131, 171

\bibitem[Kokubo \& Ida(2000)]{kok00} Kokubo, E., \& Ida, S. 2000,
Formation of protoplanets from planetesimals in the Solar nebula , Icarus, 143, 15

\bibitem[Kominami \& Ida(2002)]{kom02} Kominami, J., \& Ida, S., 2002,
The effect of tidal interaction with a gas disk on formation of terrestrial planets, Icarus, 157, 43

\bibitem[Kominami \& Ida(2004)]{kom04} Kominami, J., \& Ida, S., 2004,
Formation of terrestrial planets in a dissipating gas disk with Jupiter and Saturn, Icarus, 167, 231

\bibitem[Kominami et al.(2005)]{kom05} 
Kominami, J., Tanaka, H., \& Ida, S., 2005,
Orbital evolution and accretion of protoplanets tidally interacting 
with a gas disk I. Effects of Interaction with Planetesimals and Other Protoplanets. Icarus, 178, 540-552

\bibitem[Daisaka et al.(2006)]{kom06} 
Daisaka, K. J., Tanaka, H., \& Ida, S., 2006,
Orbital evolution and accretion of protoplanets tidally interacting 
with a gas disk II. Solid surface density evolution with type-I 
migration, Icarus, in press

\bibitem[Laughlin et al.(2004)]{lau04}
Laughlin, G., Steinacker, A., \& Adams, F. C. 2004, 
Type I planetary migration with MHD turbulence,
\apj, 608, 489

\bibitem[Levison \& Agnor(2003)]{lev03} Levison, H. F., \& Agnor, C., 2003
The role of giant planets in terrestrial planet formation, \aj, 125, 2692

\bibitem[Lissauer(1987)]{lis87} Lissauer, J. J., 1987, 
Timescales for planetary accretion and the structure of the protoplanetary disk, Icarus, 69, 249

\bibitem[McNeil et al.(2005)]{mcn05} McNeil, D., Duncan, M., \& Levison, H. F., 2005,
Effects of type I migration on terrestrial planet formation, \aj, 130, 2884

\bibitem[Nadya et al.(2006)]{nadya06} 
Nadya, G. et al., 2006,
Spitzer 24 $\mu$ survey of debris disks in the Pleiades, \apj, 649, 1028-1042.

\bibitem[Nagasawa et al.(2005)]{nag05} Nagasawa, M., Lin, D. N. C., \& Thommes, E., 2005,
Dynamical shake-up of planetary systems. I. Embryo trapping and induced collisions by the sweeping secular resonance and embryo-disk tidal interaction, \apj, 635, 578 

\bibitem[Nelson \& Papaloizou(2004)]{nel04} 
Nelson, R. P., \& Papaloizou, J. C. B., 2004,
The interaction of giant planets with a disc with MHD turbulence - IV. Migration rates of embedded protoplanets, \mnras, 350, 849

\bibitem[Nelson(2005)]{nel05} Nelson, R. P., 2005,
On the orbital evolution of low mass protoplanets in turbulent, magnetized disks, \aap, 443, 1067

\bibitem[O'Brien et al.(2006)]{obr06} O'Brien, D. P., Morbidelli, A., \& Levison, H. F., 2006,
Simulations of terrestrial planet formation with strong dynamical friction: Implications for the origin of the Earth's water, \aj, in press

\bibitem[Papaloizou \& Larwwod(2000)]{Pap00} 
Papaloizou, J. C. B. \& Larwood, J. D., 2000,
On the orbital evolution and growth of protoplanets embedded in a gaseous disc,
\mnras, 530, 823-833.

\bibitem[Raymond et al.(2006)]{ray06} Raymond, S. N., Barnes, R., \& Kaib, N. A., 2006,
Predicting planets in known extrasolar planetary systems. III. Forming terrestrial planets, \apj, 644, 1223

\bibitem[Rice \& Armitage(2003)]{ric03} Rice, W. K. M., Armitage, P J., 2003,
On the formation timescale and core masses of gas giant planets, \apj, 598, 55

\bibitem[Sano et al.(2000)]{san00} Sano, T., Miyama, S. M., Umebayashi, T., \& Nakano, T., 2000,
Magnetorotational instability in protoplanetary disks. II. Ionization state and unstable regions, \apj, 543, 486

\bibitem[Tanaka \& Ward(2002)]{tan02} Tanaka, H., \& Ward, W. R., 2002,
Three-dimensional interaction between a planet and an isothermal gaseous disk. 
I. Corotation and Lindblad torques and planet migration, \apj, 565, 1257

\bibitem[Tanaka \& Ward(2004)]{tan04} Tanaka, H., \& Ward, W. R., 2004,
Three-dimensional interaction between a planet and an isothermal gaseous disk. 
II. Eccentricity Waves and Bending Waves, \apj, 602, 388

\bibitem[Ward(1986)]{war86} Ward, W. R., 1986,
Density waves in the solar nebula - Differential Lindblad torque, Icarus, 67, 164

\bibitem[Ward(1997)]{war97} Ward, W. R., 1997,
Survival of planetary systems, \apj, 482, L211

\bibitem[Ward(1993)]{war93} Ward, W. R., 1993
Density waves in the solar nebula - Planetesimal velocities, Icarus, 106, 274

\bibitem[Yin and Ozima(2003)]{yin03} Yin, Q., Ozima, M.\
2003.\ Hf-W and I-Xe Ages and the Planetary Formation Timescale.\
Geochemica et Cosmochemica Acta Supplement 67, 564.

\bibitem[Yin et al.(2002)]{yin02} Yin, Q., Jacobsen, S.~B.,
Yamashita, K., Blichert-Toft, J., T{\'e}louk, P., Albar{\`e}de, F.\ 2002.\
A short timescale for terrestrial planet formation from Hf-W chronometry of
meteorites.\ Nature 418, 949-952.

\bibitem[Yoshinaga et al.(1999)]{yos99} Yoshinaga, K., Kokubo, E., \& Makino, J., 1999,
The stability of protoplanet systems, Icarus, 139, 328


\end{thebibliography}
\end{document}